\documentclass[reprint,prd,letterpaper,superscriptaddress,amsmath,amssymb,aps] {revtex4-1}

\usepackage{graphicx}
\usepackage{dcolumn}
\usepackage{todonotes} 
\usepackage{amsmath}
\usepackage{tabu}
\usepackage{bm}
\usepackage{siunitx}
\usepackage{caption}
\usepackage{subcaption}
\usepackage{pifont}
\usepackage{xcolor}

\usepackage[implicit=true, debug=false, 
plainpages=false, colorlinks=true, 
linkcolor=black, urlcolor=black, citecolor=black, filecolor=black, 
hyperindex=true]{hyperref}
\newcommand{\atUC}{\affiliation{Dept. of Physics, Enrico Fermi Inst., Kavli Inst. for Cosmological Physics, Univ. of Chicago, Chicago, IL 60637.}}
\newcommand{\atUCLA}{\affiliation{Dept. of Physics and Astronomy, Univ. of California, Los Angeles, Los Angeles, CA 90095.}}
\newcommand{\atOSU}{\affiliation{Dept. of Physics, Center for Cosmology and AstroParticle Physics, Ohio State Univ., Columbus, OH 43210.}}
\newcommand{\atUH}{\affiliation{Dept. of Physics and Astronomy, Univ. of Hawaii, Manoa, HI 96822.}}
\newcommand{\atNTU}{\affiliation{Dept. of Physics, Grad. Inst. of Astrophys.,\& Leung Center for Cosmology and Particle Astrophysics, National Taiwan University, Taipei, Taiwan.}}
\newcommand{\atKU}{\affiliation{Dept. of Physics and Astronomy, Univ. of Kansas, Lawrence, KS 66045.}}
\newcommand{\atWU}{\affiliation{Dept. of Physics, McDonnell Center for the Space Sciences, Washington Univ. in St. Louis, MO 63130.}}
\newcommand{\atSLAC}{\affiliation{SLAC National Accelerator Laboratory, Menlo Park, CA, 94025.}}
\newcommand{\atUD}{\affiliation{Dept. of Physics, Univ. of Delaware, Newark, DE 19716.}}
\newcommand{\atUCL}{\affiliation{Dept. of Physics and Astronomy, University College London, London, United Kingdom.}}
\newcommand{\atJPL}{\affiliation{Jet Propulsion Laboratory, Pasadena, CA 91109.}}

\newcommand{\atCalPoly}{\affiliation{Physics Dept., California Polytechnic State Univ., San Luis Obispo, CA 93407.}}

\newcommand{\atUCSD}{\affiliation{Center for Astrophysics and Space Sciences, Univ. of California, San Diego, La Jolla, CA 92093.}}
\newcommand{\atMoscow}{\affiliation{Moscow Engineering Physics Institute, Moscow, Russia.}}

\newcommand{\combinedBackgroundEstimates}{$0.64^{+0.69}_{-0.45}$}
\newcommand{\pengBGEstimates}{$0.34^{+0.66}_{-0.16}$}

\begin{document} 
\bibliographystyle{apsrev4-1} 
\title{Constraints on the ultra-high energy cosmic neutrino flux\\from the fourth flight of ANITA}

\author{P.~W.~Gorham}\atUH 
\author{P.~Allison}\atOSU
\author{O.~Banerjee}\atOSU
\author{L.~Batten}\atUCL 
\author{J.~J.~Beatty}\atOSU 
\author{K.~Belov}\atJPL 
\author{D.~Z.~Besson}\atKU\atMoscow
\author{W.~R.~Binns}\atWU 
\author{V.~Bugaev}\atWU 
\author{P.~Cao}\atUD 
\author{C.~C.~Chen}\atNTU 
\author{C.~H.~Chen}\atNTU
\author{P.~Chen}\atNTU 
\author{J.~M.~Clem}\atUD 
\author{A.~Connolly}\atOSU 
\author{L.~Cremonesi}\atUCL 
\author{B.~Dailey}\atOSU 
\author{C.~Deaconu}\atUC 
\author{P.~F.~Dowkontt}\atWU 
\author{B.~D.~Fox}\atUH 
\author{J.~W.~H.~Gordon}\atOSU
\author{C.~Hast}\atSLAC
\author{B.~Hill}\atUH 
\author{S.~Y.~Hsu}\atNTU
\author{J.~J.~Huang}\atNTU
\author{K.~Hughes}\atUC\atOSU
\author{R.~Hupe}\atOSU 
\author{M.~H.~Israel}\atWU 
\author{K.~M.~Liewer}\atJPL
\author{T.~C.~Liu}\atNTU 
\author{A.~B.~Ludwig}\atUC 
\author{L.~Macchiarulo}\atUH 
\author{S.~Matsuno}\atUH 
\author{C.~Miki}\atUH 
\author{K.~Mulrey}\atUD
\author{J.~Nam}\atNTU
\author{C.~Naudet}\atJPL
\author{R.~J.~Nichol}\atUCL
\author{A.~Novikov}\atKU\atMoscow
\author{E.~Oberla}\atUC 
\author{S.~Prohira}\atOSU\atKU
\author{B.~F.~Rauch}\atWU
\author{J.~M.~Roberts}\atUH\atUCSD
\author{A.~Romero-Wolf}\atJPL
\author{B.~Rotter}\atUH
\author{J.~W.~Russell}\atUH 
\author{D.~Saltzberg}\atUCLA
\author{D.~Seckel}\atUD
\author{H.~Schoorlemmer}\atUH
\author{J.~Shiao}\atNTU
\author{S.~Stafford}\atOSU
\author{J.~Stockham}\atKU
\author{M.~Stockham}\atKU
\author{B.~Strutt}\atUCLA 
\author{M.~S.~Sutherland}\atOSU
\author{G.~S.~Varner}\atUH
\author{A.~G.~Vieregg}\atUC
\author{N.~Wang}\atUCLA
\author{S.~H.~Wang}\atNTU
\author{S.~A.~Wissel}\atCalPoly

\collaboration{ANITA Collaboration}\noaffiliation

\date{\today}

\begin{abstract}
  The ANtarctic Impulsive Transient Antenna (ANITA) NASA long-duration balloon payload completed its fourth flight in December 2016, after 28 days of flight time.  ANITA is sensitive to impulsive broadband radio emission from interactions of
  ultra-high-energy neutrinos in polar ice (Askaryan emission). 
  We present the results of two separate blind analyses searching for signals from Askaryan emission in the data from the fourth flight of ANITA.
  The more sensitive analysis, with a better expected limit, has a background estimate of \combinedBackgroundEstimates~and an analysis efficiency of 82$\pm2\%$.  The second analysis has a background estimate of \pengBGEstimates~and an analysis efficiency of 71$\pm6\%$.  
  Each analysis found one event in the signal region, consistent with the background estimate for each analysis.  
  The resulting limit further tightens the constraints on the diffuse flux of ultra-high-energy neutrinos at energies above $10^{19.5}$~eV.

\end{abstract}

\maketitle

\section{Introduction}

Ultra-high-energy (UHE) neutrinos are expected to carry information about the highest-energy cosmic rays and their accelerators, including UHE cosmic ray (UHECR) composition and the type and cosmological evolution of their sources. 
These neutrinos may be produced directly in sources or produced as UHECRs propagate through the universe and interact with cosmological or astrophysical photon backgrounds. 
Of particular interest are cosmogenic neutrinos~\cite{bz,stecker} produced as a result of protons interacting with CMB photons through the GZK process~\cite{GZK1,GZK2}.
These neutrinos are produced within 100 Mpc of the UHECR source, and are tightly aligned to the source direction on the sky, unaffected by magnetic fields during propagation.

Detecting these UHE neutrinos is a challenge that requires instrumenting and monitoring immense volumes of dense material because the expected neutrino flux is low and the neutrino-nucleon cross section is small.  
The radio technique takes advantage of the Askaryan effect~\cite{askaryan} and the long attenuation lengths of ice to observe large volumes with minimal instrumentation.  
Coherent Cherenkov emission arises from a neutrino induced shower in a dense dielectric medium for wavelengths longer than the lateral extent of the shower.  
The expected Askaryan signal is broadband at frequencies less than a few GHz and the power emitted scales with the square of the energy of the electromagnetic shower, as observed in accelerator experiments~\cite{slac_silica,slac_ice}.

The ANtarctic Impulsive Transient Antenna (ANITA) is a NASA long-duration balloon 
payload~\cite{instrument} designed to search for broadband, impulsive radio emission from neutrinos in the Antarctic ice, which is highly radio transparent~\cite{Barwick05}. It consists of 48 high-gain, dual-polarization antennas and flies at a height of $\sim 40$~km above the Antarctic surface.  
ANITA is sensitive to Askaryan emission from neutrino-induced showers
in ice, and can also observe geomagnetic emission from
extensive air showers (EAS) induced by cosmic rays or other particles~\cite{anita1CR,anita1Harm}.
Due  to  the  approximately  vertical  direction  of  Earth's magnetic  field  in  Antarctica,  EAS  emission  is  predominately  horizontally polarized.
Askaryan  emission  visible to ANITA is mainly vertically polarized for Standard Model cross sections, due to the ice surface Fresnel transmission coefficients and Cherenkov cone geometry.

ANITA has previously placed limits on the diffuse UHE cosmic neutrino flux using data from the first three flights~\cite{anita1,anita2,anita3}.  
In this paper, we present two analyses of data from the fourth flight of the ANITA experiment (ANITA-IV), focusing on the results of the search for UHE neutrinos via their in-ice Askaryan radio emission. The analyses presented here result in an improved upper limit on the diffuse flux of UHE neutrinos at energies greater than $10^{19.5}$~eV.

We discuss the ANITA-IV instrument and flight in Sections~\ref{sec:instrument} and~\ref{sec:flight}, respectively. The methods used in the two searches for Askaryan emission from UHE neutrinos in the ANITA-IV dataset are described in Section~\ref{sec:methods}. The results, including improved upper limits on the diffuse UHE neutrino flux at the highest energies ($E>10^{19.5}$~eV), are presented in Section~\ref{sec:results}.  We conclude in Section~\ref{sec:discussion}.

\section{The ANITA-IV Instrument}\label{sec:instrument}
    \begin{figure}
   \includegraphics[width=0.8\columnwidth]{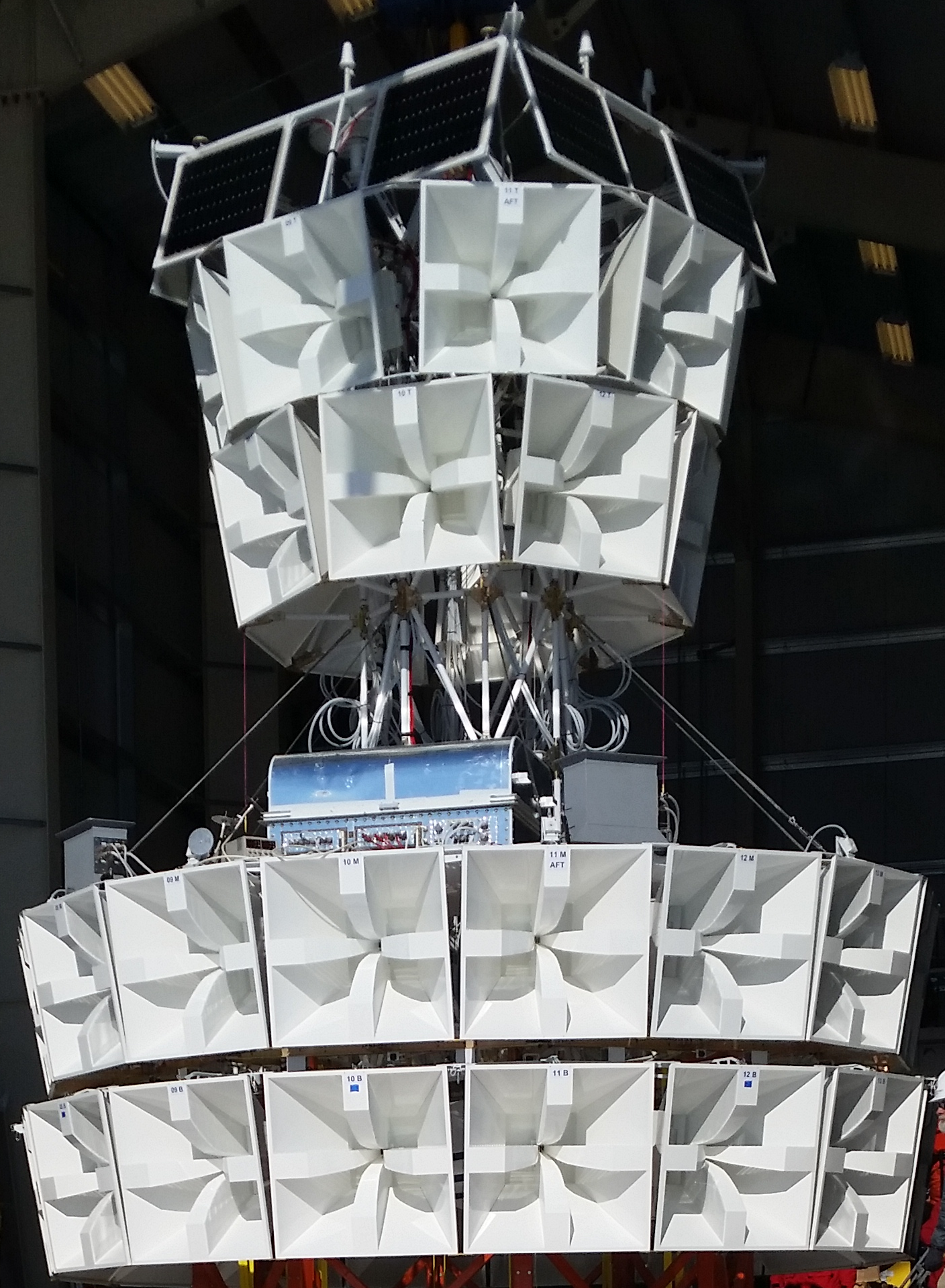} 

   \caption{A picture of the ANITA-IV payload.
   ANITA is $\sim$8~m tall, and each horn antenna is roughly 0.95~m from edge to edge.
   An additional row of photovoltaic panels (not pictured) dropped down below the bottom ring of antennas after launch.
   }
     \label{fig:a4photo}
   \end{figure}
   
 \begin{figure*}
   \includegraphics[width=0.9\textwidth]{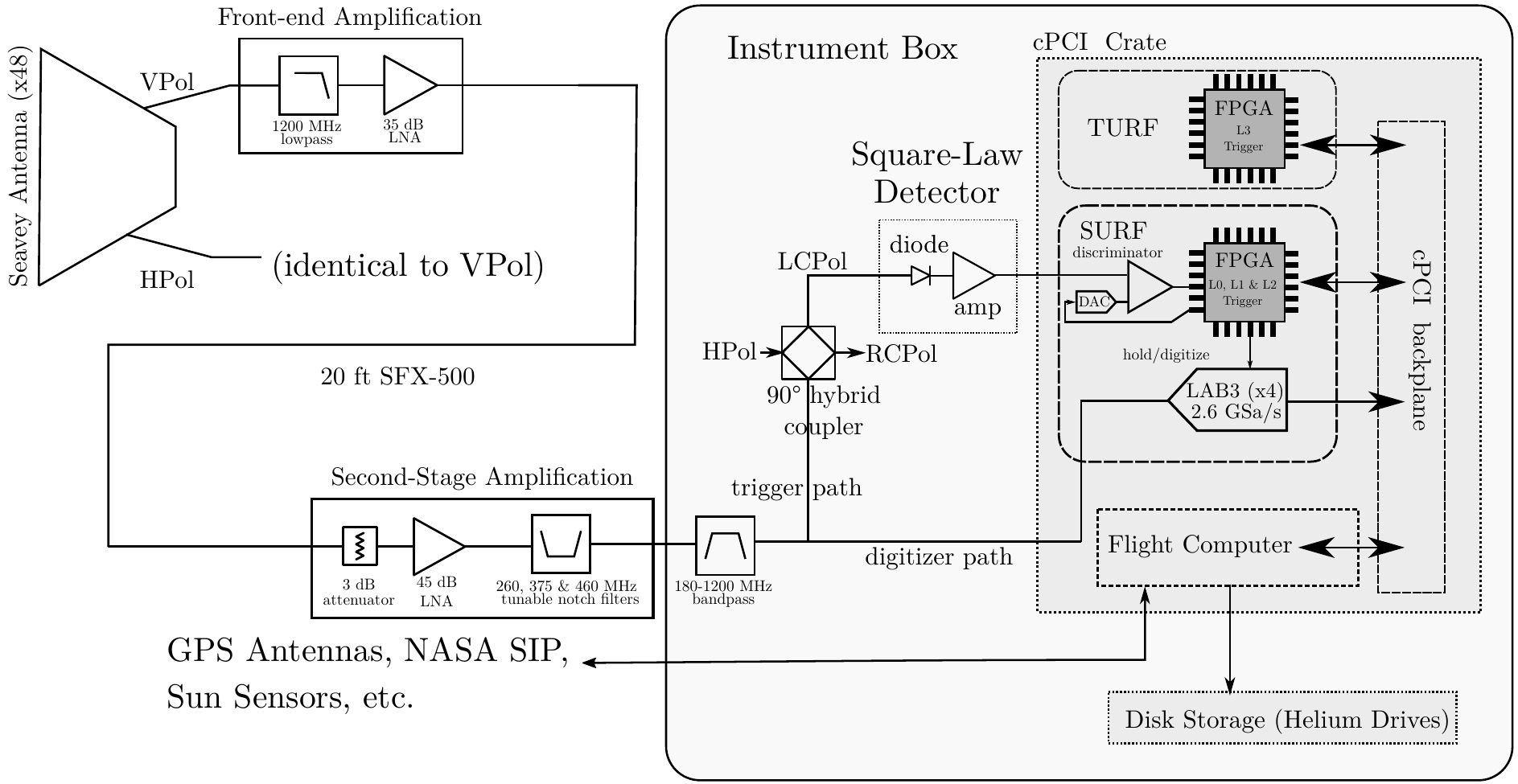} 

   \caption{A schematic diagram of the ANITA-IV instrument. See text for a full description of the electronics.
   }
   
     \label{fig:a4diagram}
   \end{figure*}

ANITA-IV retains some of the key features of 
the previous three ANITA payloads~\cite{anita1,anita2}, with significant upgrades.
The primary upgrades from ANITA-III are the addition of three tunable notch filters on each channel to better reject continuous waveform (CW) interference~\cite{TUFF} and the implementation of a trigger that requires a high fraction of linear polarization, to better reject thermal noise in favor of linearly-polarized neutrino signals. Here we briefly describe the instrument.

 Forty-eight dual-polarized quad-ridge horn antennas from Antenna Research Associates, Inc. are arranged in three rings, in a cylindrical pattern for a total of 96 broadband
  (180~MHz-1200~MHz) channels. Each ring has 16 antennas, and each antenna has azimuthally-aligned partners in each of the other two rings, forming 16 azimuthal sectors with three antennas each, as shown in Fig.~\ref{fig:a4photo}.  
   A schematic of the ANITA-IV instrument and data acquisition system is depicted in
  Fig.~\ref{fig:a4diagram}. 
  The signal from each channel is low-pass filtered and amplified by a custom-built low-noise amplifier mounted on each antenna.  After the first stage of amplification, the signal is then sent through a second stage amplifier and notch filtered and band-pass filtered before being split into trigger and
  digitization paths. Antenna temperatures are typically $\sim$130~K and the additional noise temperature from the front-end filters and amplifiers is $\sim$65~K.  
  
  The addition of notch filters on this flight was important for keeping low trigger thresholds and high digitization livetime, defined as the fraction of time ANITA has unfilled buffers and is able to record events.
  ANITA's trigger is designed to be ``threshold-riding,'' meaning we dynamically tune thresholds so that the global trigger rate is approximately 50~Hz, which is the fastest we can trigger without impacting instrument livetime.
  Notch filters were added in order to maintain this threshold-riding trigger scheme.
  
  ANITA-III recorded a significant fraction of events that contained CW interference from military satellites and Antarctic base communications systems.  
  In ANITA-III this interference was dealt with mainly by restricting portions of the payload from triggering (``masking'').  
  Because these satellites were geosynchronous and almost always in view, this meant the north-facing half of the instrument was almost always masked off, reducing ANITA-III's total neutrino acceptance~\cite{TUFF}.  
  ANITA-IV's notch filters were installed at default center frequencies of 260 MHz, 375 MHz, and 460 MHz to mitigate the effects of sources of CW noise.  
  The notches could be switched on and off, as well as tuned in frequency space in flight, depending on the noise environment of the payload.  
  As a result of these notch filters, the masked fraction of the payload was always below 30\% in ANITA-IV~\cite{TUFF}.

  The trigger path uses a 90-degree hybrid coupler that combines the horizontally and vertically-polarizated signals from each antenna, producing left circular polarization (LCP) and right circular polarization (RCP).  
  The hybrid output feeds the LCP and RCP outputs into a custom tunnel diode that acts as a fast square-law detector. 
  Each channel compares the output of a tunnel diode to a dynamically-adjusted
  threshold to determine if a channel-level (zeroth-level) trigger should be issued. 
  Zeroth-level triggers are entirely dependent on the total power in the signal.  If a zeroth-level trigger is issued, a check for a first-level trigger is initiated.
    The trigger thresholds are adjusted in real time to keep the zeroth-level trigger rate approximately at its target rate, which varied between 5~MHz and 6~MHz for ANITA-IV.
  
  A first-level trigger is only issued if both the LCP and RCP outputs exceed the required threshold within 4~ns of one another.  
  Our expected science triggers should be mainly linearly polarized, and enforcing this LCP/RCP coincidence requirement is equivalent to requiring a high fraction of linear polarization of the signal along any axis.

A second-level trigger condition is imposed at the level of each azimuthal sector and is satisfied by a coincidence of
  two or more channels.  If a first-level trigger is issued for a given channel, a 
  coincidence window opens during which another channel in the
  same azimuthal sector issuing a first-level trigger would
  generate a second-level trigger. 
  A second-level trigger begins by delaying the signal from the middle/bottom rings by 4~ns, which biases against triggering on signals where the top and middle or top and bottom issue a first-level trigger at the same time.  
  The size of each coincidence window depends on the ring that issued the first first-level trigger. The windows are set to preferentially trigger on signals coming from below the horizon: 12~ns for the bottom to top ring, 
  8~ns for the middle to top ring, 
  	and 4~ns for the bottom to middle ring.

The third-level (global) trigger is generated by the coincidence of second-level
triggers occurring in two adjacent azimuthal sectors within 10~ns of one another.  A third-level
trigger will cause the digitized signals to be read out, 
unless a four-deep digitizer buffer is full. Over the course of the flight, the average deadtime incurred from full digitizer buffers was 6.7\%.  The third-level 
trigger rate over the course of the flight for ANITA-IV was approximately 50~Hz.
  
In addition to the science triggers generated by the trigger logic described above, a set of ``minimum-bias triggers'' are also recorded. These triggers are taken on the instruction of on-board computers or pulse per second signals from the on-board GPS units and do not follow the normal trigger logic. Minimum-bias triggers are used to characterize the noise environment throughout the flight.

Triggered signals are digitized on LAB3~\cite{lab4} switched capacitor array digitizers.  Each channel samples at 2.6~GSa/s and has four 260-sample analog buffers.

ANITA has two methods for lowering trigger rates to prevent buffers from being filled and incurring deadtime.  
The previously mentioned tunable notch filters are the primary method for mitigating CW (narrow bandwidth) sources.  
Any of the three notches per channel can be tuned to a given frequency and toggled on in response to strong CW sources, determined by monitoring the frequency spectra of recently triggered events.  
There is also an ability to mask out a portion of the payload and prevent it from triggering if a certain direction is contaminated by significant noise at a given time.  
Trigger masks, when enabled, are automatically applied by the on-board computer to azimuthal sectors with second-level or third-level trigger rates that exceed some threshold.  
This dynamic masking is intended to cut out any areas of ANITA's view that contain significant sources of anthropogenic noise.  
In ANITA-IV, masking was enabled near the McMurdo and South Pole stations.

\section{The ANITA-IV Flight}\label{sec:flight}

\begin{figure}
   \includegraphics[width=\linewidth]{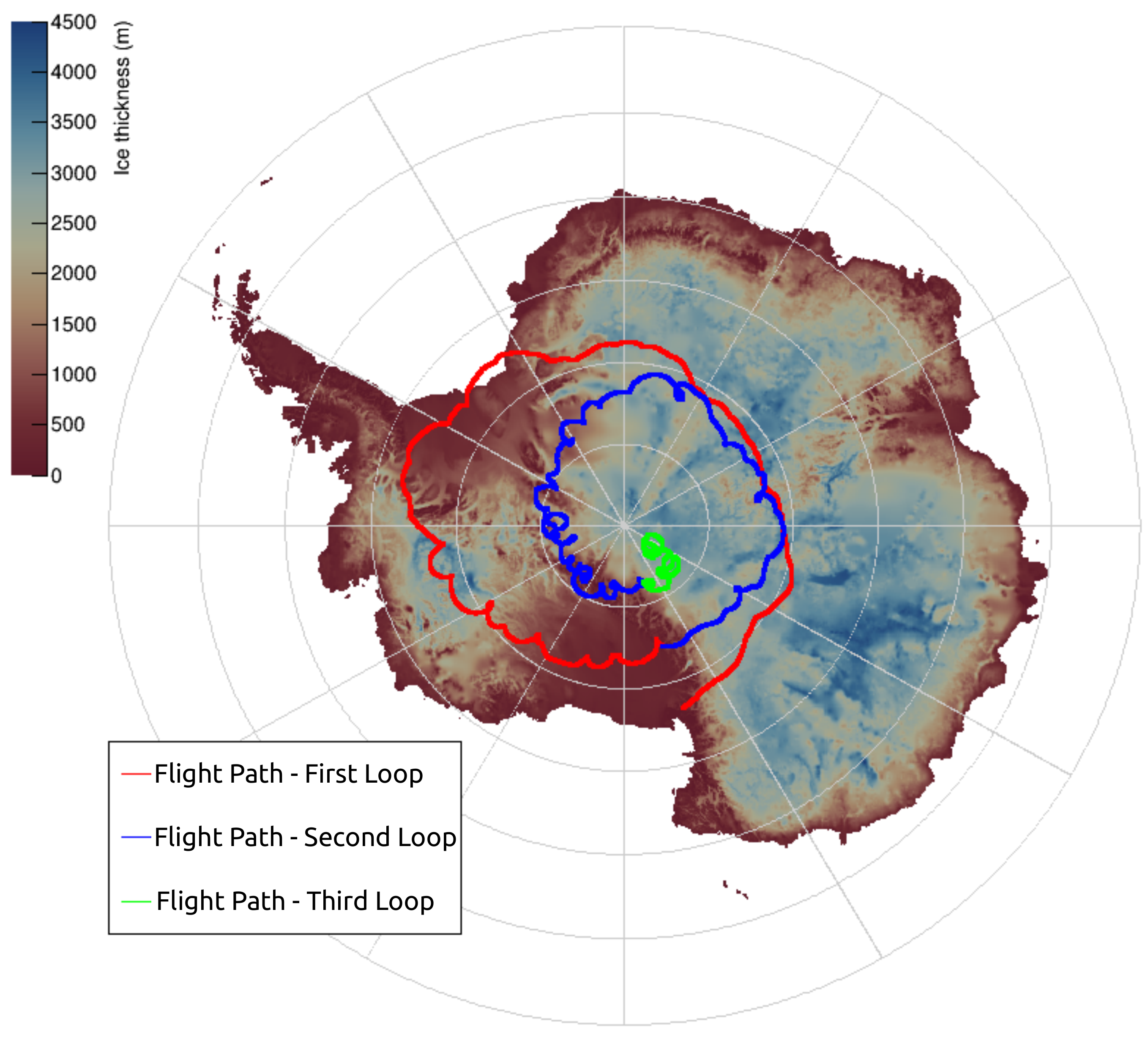} 

   \caption{The ANITA-IV flight path.  The flight began at McMurdo Station and made two full passes around the continent, before landing near South Pole Station.}
     \label{fig:flightPath}
   \end{figure}

  ANITA-IV launched from the NASA Long-Duration Balloon facility on the Ross Ice
  Shelf near McMurdo Station on December $2^{\textrm{nd}}$, 2016.  ANITA-IV flew for
  28 days before termination on December $29^{\textrm{th}}$, 2016.  The flight path is shown in Fig.~\ref{fig:flightPath}.  
  The hard disks and flight hardware were recovered from their landing site near South Pole Station. 

 Impulsive calibration signals were 
 sent to ANITA-IV from high-voltage calibration pulsers deployed at the
  launch site and at
  WAIS Divide, in West Antarctica. The ANITA team transmitted calibration pulses with horizontal, vertical, and 45-degree polarization.  The WAIS Divide pulsers are referenced to GPS time
  to facilitate identification.  WAIS pulser data was used for calibration of antenna phase centers, as well as relative horizontal and vertical polarization channel timings.  
  
  In addition to calibration pulses from WAIS, the High-Altitude Calibration (HiCal)-2 instrument flew as a companion balloon to the ANITA-IV experiment~\cite{hical2_instrument}.  
  The HiCal instrument is a balloon-borne broadband calibration pulser that follows ANITA.  HiCal-2 comprised two payloads, which flew a combined 18 days.  
  ANITA detected over 10,000 pulses from the HiCal payloads, both direct signals from the HiCal itself, and companion signals that were reflected off of the surface of the ice before being recorded by ANITA.  Pairs of direct and reflected pulses provide measurements of Antarctic ice surface Fresnel coefficients at a variety of angles important for both analysis and simulation~\cite{hical2_reflectivity}.


\section{ANITA-IV Analysis}\label{sec:methods}

ANITA-IV recorded over ninety million triggers throughout its flight.
The trigger on the instrument is set so that the vast majority ($\sim99\%$) of
those events are thermal noise, the level of which dictates ANITA's threshold.  
The majority of the remaining events are anthropogenic transient and CW emission and occasional impulsive emission believed to be electromagnetic interference that escapes our Faraday enclosure, which we call {\it payload blasts}.
 
After reviewing the backgrounds to the search and the simulation tools, we will
briefly summarize both of the searches performed for neutrino-induced Askaryan emission in ice.  
Much of the analysis is based on techniques used in the ANITA-II and ANITA-III Askaryan neutrino searches \cite{anita2,anita3}.

\subsection{Backgrounds}

Random fluctuations of thermal noise from the combination of ice and sky in the field of view of the antennas account for most of the recorded ANITA-IV events.
In order to maximize sensitivity, ANITA's trigger threshold is set so that the global trigger rate is approximately 50~Hz throughout the whole flight. 
Because neutrino and anthropogenic sources are rare relative to thermal noise sources, we are dominated by thermal noise triggers.

Triggered events from anthropogenic CW from terrestrial 
transmitters or satellites have been greatly reduced in ANITA-IV as a result of the tunable notch filters.  Because of this, most of the anthropogenic triggers are broadband in nature.

Payload blasts are impulsive radio-frequency emission whose source remains unknown, although are consistent with being generated by electronics on the ANITA payload. 
Payload blasts are characterized by non-planar wavefront geometry, a distinct, low frequency dominated spectrum, and are typically much
stronger in the bottom and middle rings of antennas than the top ring, indicative of an origin local to the payload.

Thermal noise fluctuations that by chance appear impulsive and reconstruct as coming from an isolated part of the continent, and isolated, broadband, impulsive anthropogenic emission from the ground are both sources of background that remain after analysis cuts are developed, although the latter dominates the background that remains in the signal region after all cuts are applied. 
These two sources of background are estimated, with systematic uncertainties, before unblinding each analysis.

\subsection{Simulation} 

ANITA's primary simulation tool is \texttt{icemc}, described in greater detail
in~\cite{icemc}. 
The \texttt{icemc} program fully simulates the ANITA trigger
and digitizer signal chains and uses the flight paths and recorded channel
thresholds in order to model the acceptance of ANITA.  
It is a weighted Monte Carlo (MC) simulation where each neutrino is generated along with a weight
accounting for earth absorption and a phase-space factor. 

Our efficiency is calculated using a set of neutrinos generated by \texttt{icemc}.  The simulated neutrinos follow the maximum mixed-composition Kotera {\it et al.}~\cite{kotera} flux
model, with Standard Model cross sections~\cite{crosssection}. Most neutrino observables are model independent, while changes to neutrino cross sections can result in different distributions for expected observed angle and polarization angle.  Noise is added to the digitizer and trigger paths separately.  Noise in the trigger path is generated from distributions created from minimum-bias triggers taken throughout the flight.  The digitizer path adds simulated neutrino waveforms to real, stored minimum-bias triggers.  

\subsection{Searches for Askaryan emission from UHE neutrinos}

   \begin{figure*}
  \centering
  \begin{minipage}{\columnwidth}
    \includegraphics[width=\columnwidth]{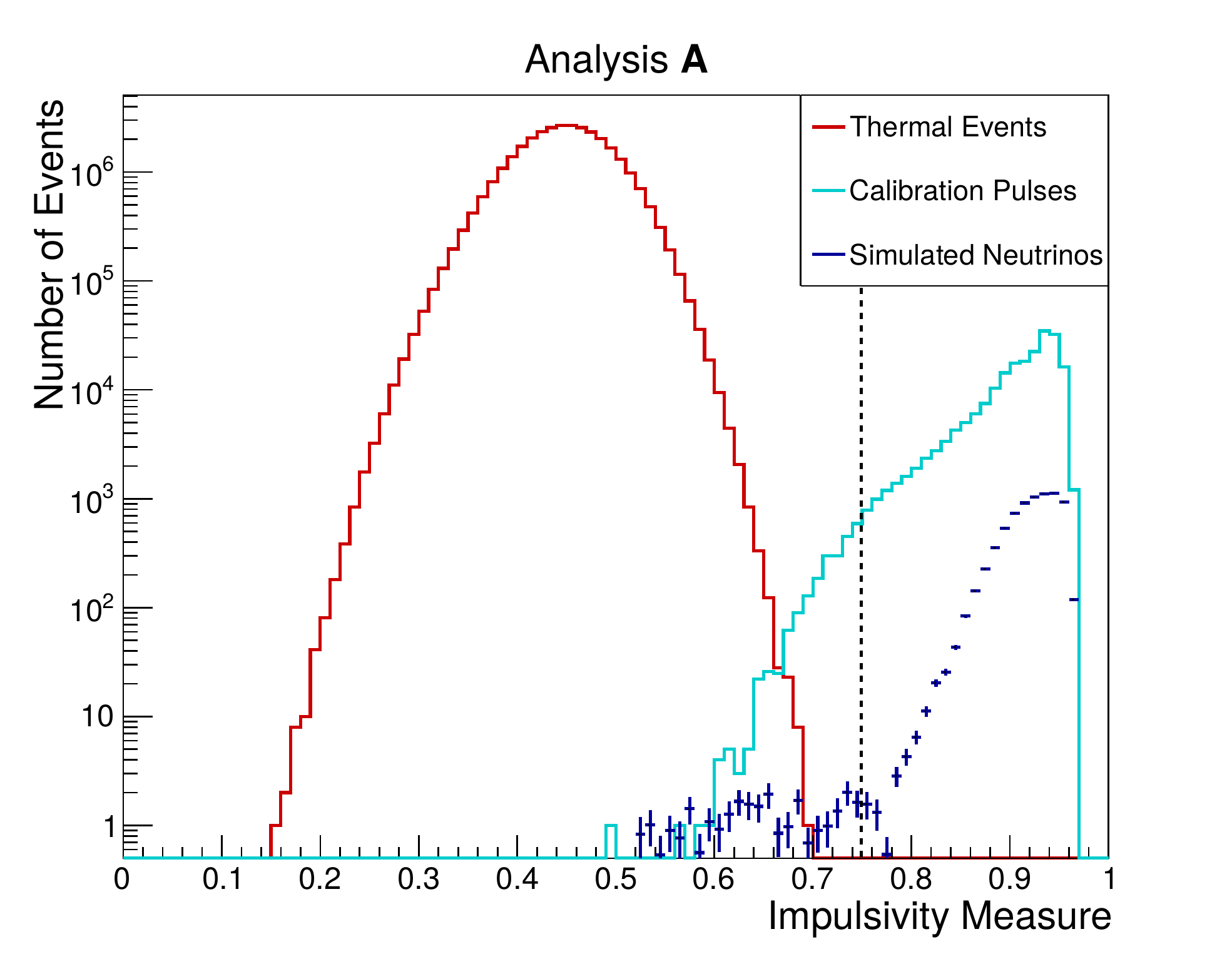}
  \end{minipage}
  \hfill
  \begin{minipage}{\columnwidth}
    \includegraphics[width=\columnwidth]{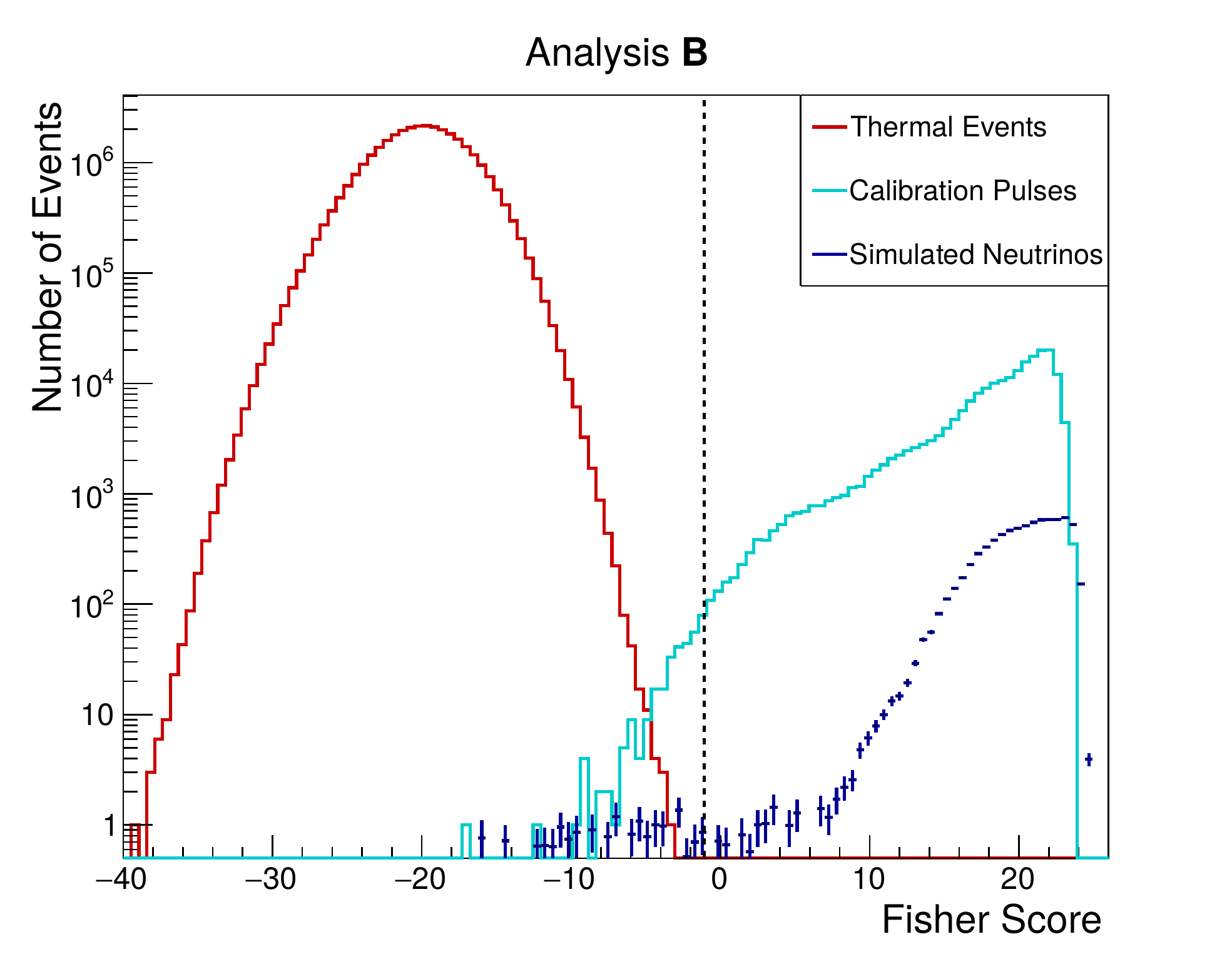}
  \end{minipage}
  
    \caption{Primary discriminator between signal and thermal noise for Analysis \textbf{A} (left) and \textbf{B} (right).  Analysis \textbf{A} uses the impulsivity of the signal, which runs from 0 (non-impulsive) to 1 (a cut value of 0.75 was chosen, shown as a dashed line).  Analysis \textbf{B} uses a multi-variate Fisher discriminant that includes impulsivity as one of its input variables (a cut value of -1 was chosen, shown as a dashed line).}
  \label{fig:analysis_metrics}
\end{figure*}

Two independent Askaryan neutrino analyses were performed, which we denote,
in order of completion, \textbf{A} and \textbf{B}. Analyses \textbf{A} and \textbf{B} are similar to each other and to previous ANITA analyses in using common criteria across the continent and searching for isolated events~\cite{anita1,anita2,anita3}.  Both neutrino searches were blind, with the region of parameter space where the signal resides kept hidden until cuts were established, and analysis efficiencies and background estimates calculated.

Despite the addition of tunable notch filters in the instrument, there is still a need to filter waveforms to mitigate undesired CW
contamination that would otherwise interfere with the analysis.  Both analyses begin with this step, using an adaptive time-domain phasor removal technique~\cite{anita3}.

The filtered waveforms from antennas with at least a partial common field of view (usually 15 antennas) are correlated against each other to produce an
interferometric map~\cite{interferometric}, which indicates the average correlation between pairs of antennas as a function of incoming direction. 
The three largest peaks in each map are considered coherent source hypotheses, and coherently summed waveforms are produced accounting for expected time delays for a signal from that direction. Additionally, we remove the group delay of the instrument response individually from each channel before coherently summing, and produce dedispersed, coherently summed waveforms for each source hypothesis.

The raw waveforms, interferometric map, and coherent waveforms, are then used to compute observables that define cuts to select a signal region.  
Examples of observables include the peak correlation value of the interferometric map, measures of coherent and dedispersed waveform impulsivity~\cite{anita3},  and polarimetric quantities.  
Both searches have their own set of ``quality cuts'' used to remove digitizer glitches ($\sim$1\% of events), payload blasts, and other poor-quality events.
 Analysis~\textbf{A} also included a cut on events whose reconstructed direction was to an azimuthal sector that was trigger-masked at the time.

Analyses \textbf{A} and \textbf{B} use similar approaches to reject thermal (non-impulsive) noise. 
Analysis \textbf{A} uses the dedispersed waveform's impulsivity measure to distinguish signal from non-impulsive (thermal) background.  
Analysis \textbf{B} uses a multivariate linear discriminant (Fisher discriminant~\cite{fisher}) on various observables, including separate measurements of impulsivity and linear polarization content, to discriminate signal-like events from non-impulsive background events. 
This discriminant is trained with simulated events as a signal sample and events reconstructing above the horizontal as a background-only sideband region (a region of phase space adjacent to the neutrino signal region that is useful for determining cut values and estimating efficiencies and backgrounds).
We use as the background sample events whose direction reconstructs to the angular region above the horizontal because we expect them to be representative of a thermal-noise-like sample: they are non-impulsive due both to a lack of impulsive sources within the atmosphere (above the payload) and the dispersive nature of the ionosphere.
Fig.~\ref{fig:analysis_metrics} shows the distribution of values given by these two metrics on simulated neutrinos and the thermal-noise-like sideband.  Events are required in both analyses to reconstruct to the continent in order to be considered further in the Askaryan neutrino search.

Events passing the signal selection that point to the continent are then grouped together in clusters based on where they originated on the continent in order to separate isolated signal-like events from anthropogenic events, which tend to cluster with each other and with known locations of human activity. 
Analyses~\textbf{A} and~\textbf{B} project a two-dimensional Gaussian distribution corresponding to the pointing resolution in azimuth and elevation for each passing
event onto a map of Antarctica, creating an event localization distribution on the continent. Analysis~\textbf{B} considers the overlap of each event's localization with the sum of the localizations of all other events, applying a cut on the angular distance between events, characterized by a log-likelihood.  Analysis~\textbf{A} instead considers whether the projected localization distribution overlaps with any single other event's projected localization distribution.
Analysis~\textbf{A} also includes \textit{a priori} information about anthropogenic sources by comparing the projected error ellipse to known areas of human activity.  
Analysis~\textbf{B} additionally considers how close each event is to the nearest event that also passes signal-like cuts, where a fit along the continent's surface is used to find the best mutual location for each event pair by assuming they came from the same location, and placing a cut at a distance of 40~km.  

Both searches treat all events the same way, regardless of polarization, but only primarily vertically-polarized events that pass all cuts are in the Askaryan neutrino signal region.  
Horizontally-polarized events that pass all cuts contain a sample of EAS.  
Both analyses also include horizontally-polarized events that reconstruct above the horizon but below the horizontal as viewed by ANITA in the sample of events that contain EAS candidates.
In Analysis~\textbf{B}, additional cuts based on the known characteristics of EAS events from previous ANITA flights are then applied to the horizontally-polarized region, to enrich the purity of the EAS sample.  

Analyses \textbf{A} and \textbf{B} set their final thermal and clustering cuts to optimize for sensitivity of the Askaryan search on the Kotera SFR1 flux model~\cite{kotera}. 
Analysis~\textbf{A} estimates backgrounds with sidebands as in the on-off problem~\cite{LiMa,RolkeOnOff}, without asserting a model for the background distributions. 
Analysis~\textbf{B} also uses an on-off treatment for the anthropogenic background, using an empirical model for the background distributions. 
In both cases, the thermal and anthropogenic backgrounds are estimated separately.
Events that reconstruct above the horizontal are used to estimate the thermal background leakage from the impulsivity measure (\textbf{A}) or multivariate discriminant (\textbf{B}). 

To estimate the anthropogenic background, Analysis~\textbf{A} uses
sidebands of very small clusters (2-6 events) and single events (called {\it singlets}) in known locations of human activity to estimate the background in the signal region, which is
singlets in locations away from human activity.

Analysis~\textbf{B} considers events from small clusters (2-100 events) to be in the sideband region.
The cutoff on cluster size was determined by taking subsamples of calibration pulser events and finding the number of events required to have all of the events in a subsample cluster together without leakage.  Analysis~\textbf{B}
estimates the anthropogenic background in the signal region using two independent variables: whether the event has a linear polarization fraction consistent with simulation and whether it is in a small cluster or is a single, isolated event. 
The signal region is single events that have a sufficiently high linear polarization fraction.

Both analyses use a profile-likelihood method~\cite{Rolke} implemented via the RooStats framework~\cite{RooStats} to optimize their final thermal and clustering cuts.
Analysis \textbf{A} has a
total estimated background in the Askaryan signal region of $0.34^{+0.49}_{-0.26}$~events and
Analysis~\textbf{B} expects \combinedBackgroundEstimates~background events.
The uncertainty on the background estimates is from the statistical uncertainty due to the sideband sample size.  
Analysis~\textbf{B} combines this with a systematic error determined by varying the sizes of clusters allowed to contribute to the background estimate calculation.

The total analysis efficiency after all cuts are applied, estimated for the Kotera flux model~\cite{kotera} using simulation, is 71$\pm6\%$ for
Analysis~\textbf{A} and 82$\pm2\%$ for Analysis~\textbf{B}.  
The systematic uncertainty on the analysis efficiency is estimated for both analyses by comparing the calculated efficiency on calibration pulser events to the efficiency on simulated neutrino events.
Statistically, Analysis~\textbf{B} is the more sensitive analysis (produces the best expected upper limit in the no-signal hypothesis).
The analysis efficiency on calibration pulser events, through the thermal-noise-like cut stage but not including clustering efficiency, is $\sim$97\% for both analyses.

\section{Results}\label{sec:results}
   \begin{figure} 
   \includegraphics[width=\columnwidth]{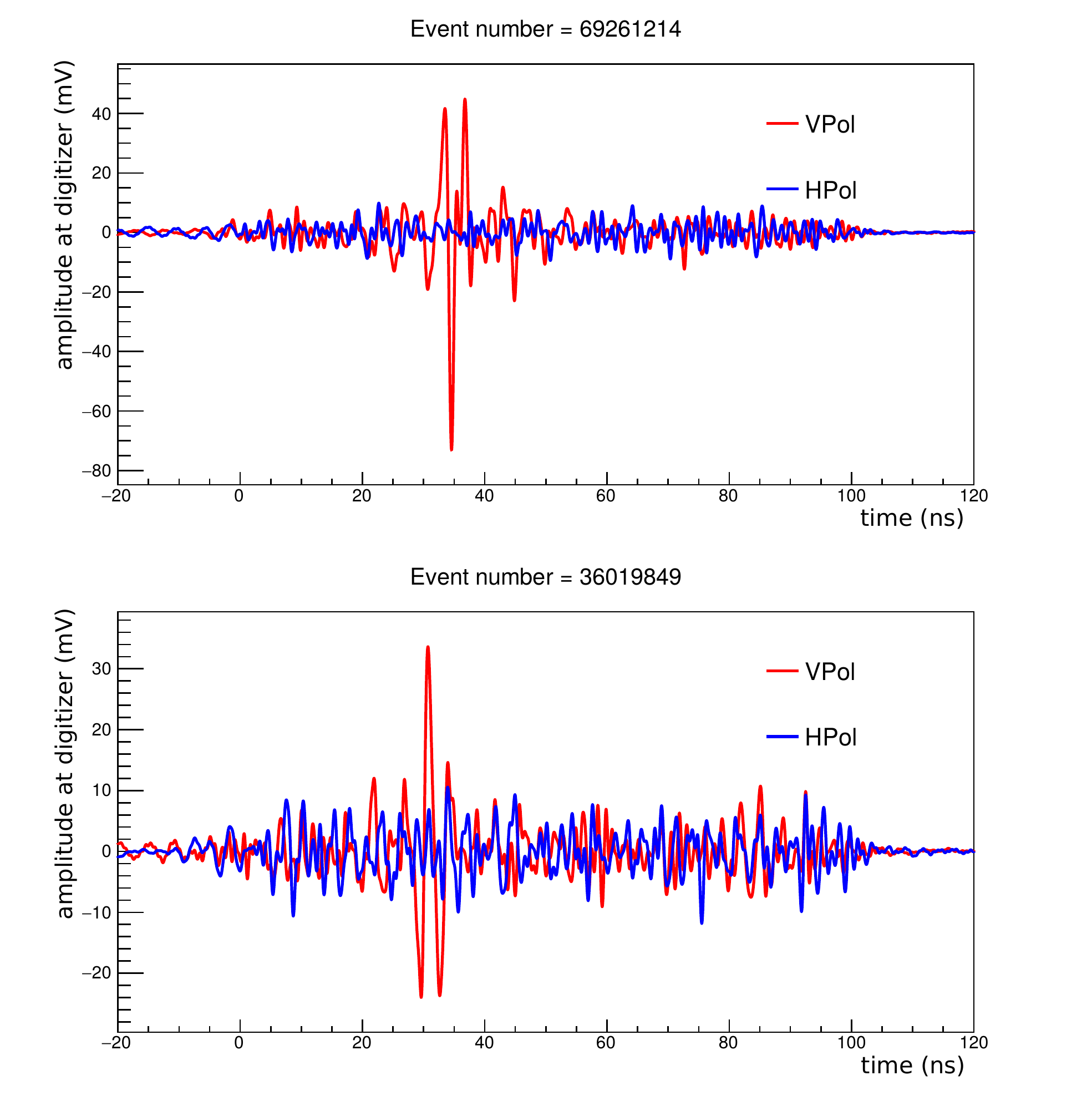} 

   \caption{Top Panel: Event 69261214, the single vertically-polarized candidate from Analysis~\textbf{A}.  Bottom Panel: Event 36019849, the single vertically-polarized candidate from Analysis~\textbf{B}.  The displayed events have the instrument response dedispersed.}
     \label{fig:both_events}
   \end{figure}
Askaryan neutrino signals are
expected to be predominantly vertically polarized for Standard Model cross sections. As such, horizontally-polarized events are not in the Askaryan neutrino signal region, but they provide a useful
cross-check on the analyses. Within the horizontally-polarized sideband region are any EAS events from cosmic
rays as well as a sub-class of cosmic-ray-like events with opposite polarity compared to EASs induced by cosmic rays, as found in ANITA-I and ANITA-III~\cite{mysteryEvent, anita3me}.  Further analysis characterizing EAS events will be published separately. 

\subsection{Summary of events found} 
 \label{sec:summaryEvents}
\begin{table*}[t]
\centering

\begin{tabular*}{.7\textwidth}{@{\extracolsep{\fill}} | l | l | l | l | l | }
\hline
Cut Name & Analysis \textbf{A} VPol & Analysis \textbf{A} & Analysis \textbf{B} VPol & Analysis \textbf{B}\\
& Events Remaining & MC Efficiency & Events Remaining & MC Efficiency\\
\hline
None & 52242901 & 1.0 & 52242901 & 1.0 \\
Quality & 17718942 & 0.919 & 37408254 & 0.981 \\
Thermal & 409455 & 0.905 & 575067 & 0.978 \\
Clustering & 1 & 0.707 & 1 & 0.819 \\
\hline
\end{tabular*}

\caption{ Summary of the effect of cuts in analysis \textbf{A} and \textbf{B}.  The quoted efficiency is the cumulative efficiency on Monte Carlo neutrinos of all cuts when performed in sequence.  Quality cuts includes removing things like digitizer glitches and ``payload blasts."  Thermal cuts removes non-impulsive events.  Clustering removes impulsive anthropogenic events.  Both analyses are left with one event in the signal region at the end of all of these cuts.} 

\label{tbl:events}
\end{table*}

Analysis~\textbf{A} finds one event in the
Askaryan signal region (Event 69261214, shown in Fig.~\ref{fig:both_events}, top panel) and 26 events in the horizontally-polarized region that contains EAS events (including above-horizon but below-horizontal events as viewed by ANITA).  
One of the 26 events is consistent with a payload blast event, and the remainder are consistent with EASs.  
Event 69261214 was cut from Analysis~\textbf{B} because it was within 40~km of other events, a requirement that Analysis~\textbf{A} did not impose, instead relying solely on a more aggressive event localization overlap cut based on the pointing resolution. 
This event passes all other Analysis~\textbf{B} cuts.  
It is consistent with the background estimate of \pengBGEstimates.
   
Analysis~\textbf{B} identifies one event in the Askaryan neutrino signal region (Event 36019849, shown in Fig.~\ref{fig:both_events}, bottom panel) and 30 events in the horizontally-polarized region that contains EASs (including above-horizon but below-horizontal events as viewed by ANITA). 
Event 36019849 was cut from Analysis~\textbf{A} because its direction reconstructed to an azimuthal sector that was trigger-masked at the time.  
The 30 horizontally-polarized events found in Analysis~\textbf{B} include the 25 EAS-like events found in Analysis~\textbf{A} and five additional EAS-like events.  
The single vertically-polarized event in Analysis~\textbf{B} is consistent with the background estimate of \combinedBackgroundEstimates.
Table~\ref{tbl:events} details how many events were cut from the analysis at each stage, and the efficiency on MC neutrinos at each step, culminating in the single event in each analysis.
The relative numbers of EAS-like events that pass in each analysis is consistent with the analysis efficiencies estimated for each analysis.

While there is overlap in the sensitivities of the two analyses, each analysis is not completely efficient and they do explore somewhat different regions of parameter space.  
True signal events will tend to pass or fail both analyses, whereas background events
probe regions near the pass/fail boundaries; however, since the analyses were
performed independently, each of the two events remains as a candidate subject only to
the expected background for its own analysis.

\subsection{Diffuse neutrino flux limit} 
   \begin{figure} 
   \includegraphics[width=\columnwidth]{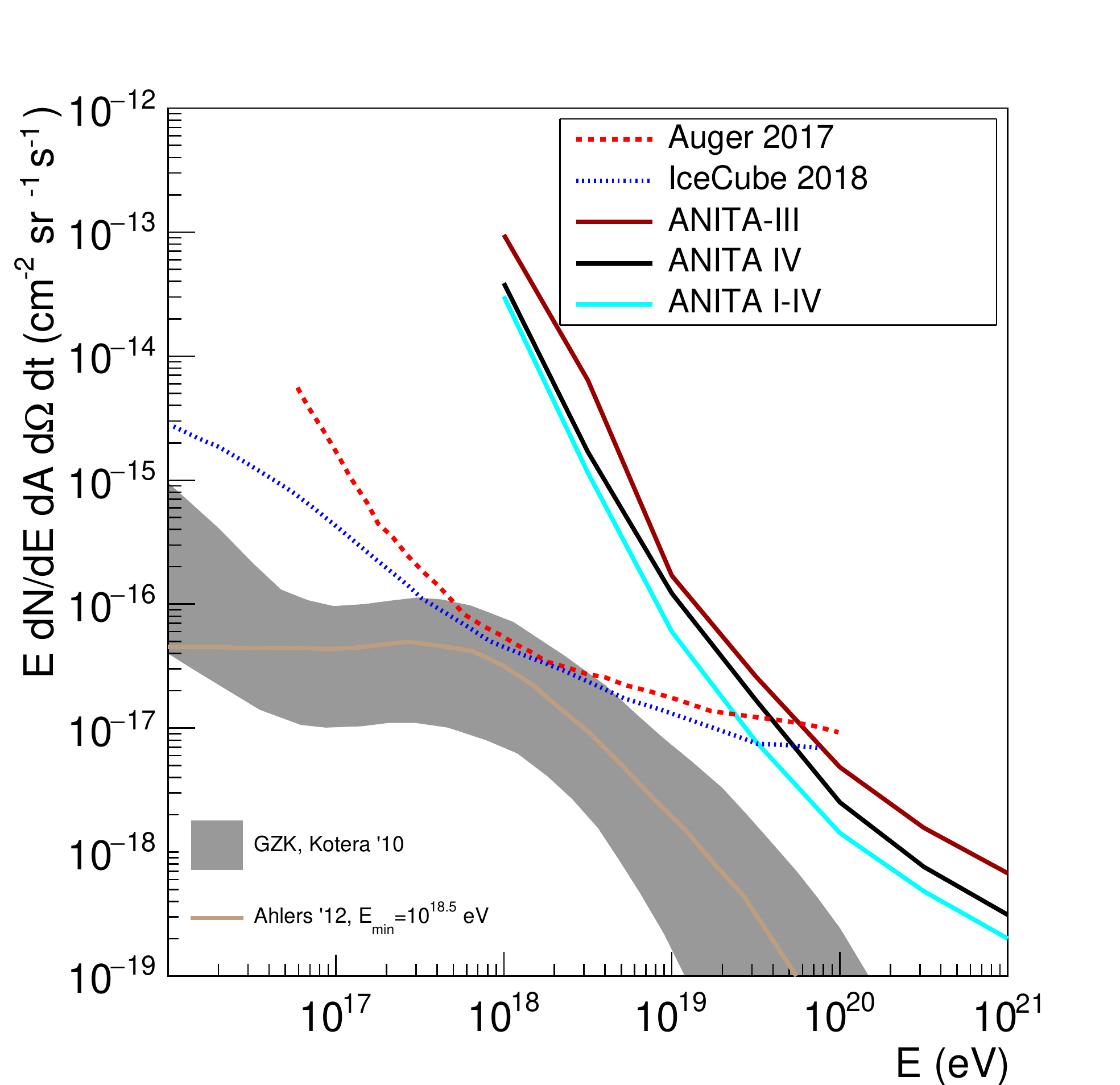} 
   
\begin{tabular} 
    {l|c|c|c|c|c|c|c}
    $\log_{10}$(E(eV)) & 18 & 18.5 & 19 & 19.5 & 20 & 20.5 & 21 \\
    \hline
  A (km$^2 \cdot$sr) & 0.0032 & 0.033 & 0.43 & 3.1 & 21 & 68 & 167\\
  \end{tabular}
  
   \caption{ ANITA-IV limit on the all flavor diffuse UHE neutrino flux and a combined limit from ANITA
I-IV.  The combined limit is made using the ANITA-IV limit shown here and the published
ANITA-I, II, and III limits~\cite{anita1,anita2,anita3}. The most recent UHE
neutrino limits from the Auger~\cite{auger2017} and IceCube\cite{icecube2018}
experiments, and two cosmogenic neutrino models~\cite{kotera,ahlers} are
also displayed.  The table lists the ANITA-IV effective area as
a function of neutrino energy used to make the limit, not
including analysis efficiency.}
     \label{fig:limit_fig}
   \end{figure}

  Fig.~\ref{fig:limit_fig} shows the neutrino flux limit, calculated using a livetime of 24.25 days and computed from the geometric mean of the acceptance computed using \texttt{icemc} and an independently developed MC simulation for ANITA, the analysis efficiency as a function of neutrino energy, and the 90\% upper limit Feldman-Cousins factor for the number of events detected and expected backgrounds. 
  Analysis~\textbf{A} and Analysis~\textbf{B} find the same number of events on similar backgrounds, but Analysis~\textbf{B} has a 10\% better expected sensitivity, so we use its result (a single observed event on a background estimate of \combinedBackgroundEstimates)~to set the limit.
   The expected number of events for a Kotera maximum all-proton and maximum mixed-composition models are
   0.33 and 0.06,
   respectively~\cite{kotera}.  
  We also set a 90\% CL integral flux limit on a pure $E_{\nu}^{-2}$ spectrum for $E_{\nu} \in [10^{18} \mathrm{eV},10^{21} \mathrm{eV}]$  of $E^2_{\nu} \Phi_{\nu} \leq  2.2 \times 10^{-7}~\mathrm{GeV}~\mathrm{cm}^{-2}~\mathrm{s}^{-1}~\mathrm{sr}^{-1}$.   
  Using the central normalization value from the IceCube best fit $E^{-\gamma}$ power law flux~\cite{icecube2017_icrc}, and assuming that the power law extends unbroken to at least $10^{20.5}$~eV, we constrain
  $\gamma$ to be greater than 1.85 at the 90\% CL using a Feldman-Cousins construction.
   
   We also show a combined ANITA~I-IV limit, where we have used the total number of events seen, the total expected background, and the sum of the effective volume for each flight weighted by their analysis efficiencies.
   The combined ANITA~I-IV upper limit improves upon previously reported limits from ANITA, and is the strongest constraint on the UHE diffuse neutrino flux between $10^{19.5}$~eV and $10^{21}$~eV to date.

\section{Discussion}\label{sec:discussion}

      \begin{figure*}
  \centering
  \begin{minipage}{.45\textwidth}
    \includegraphics[width=\columnwidth]{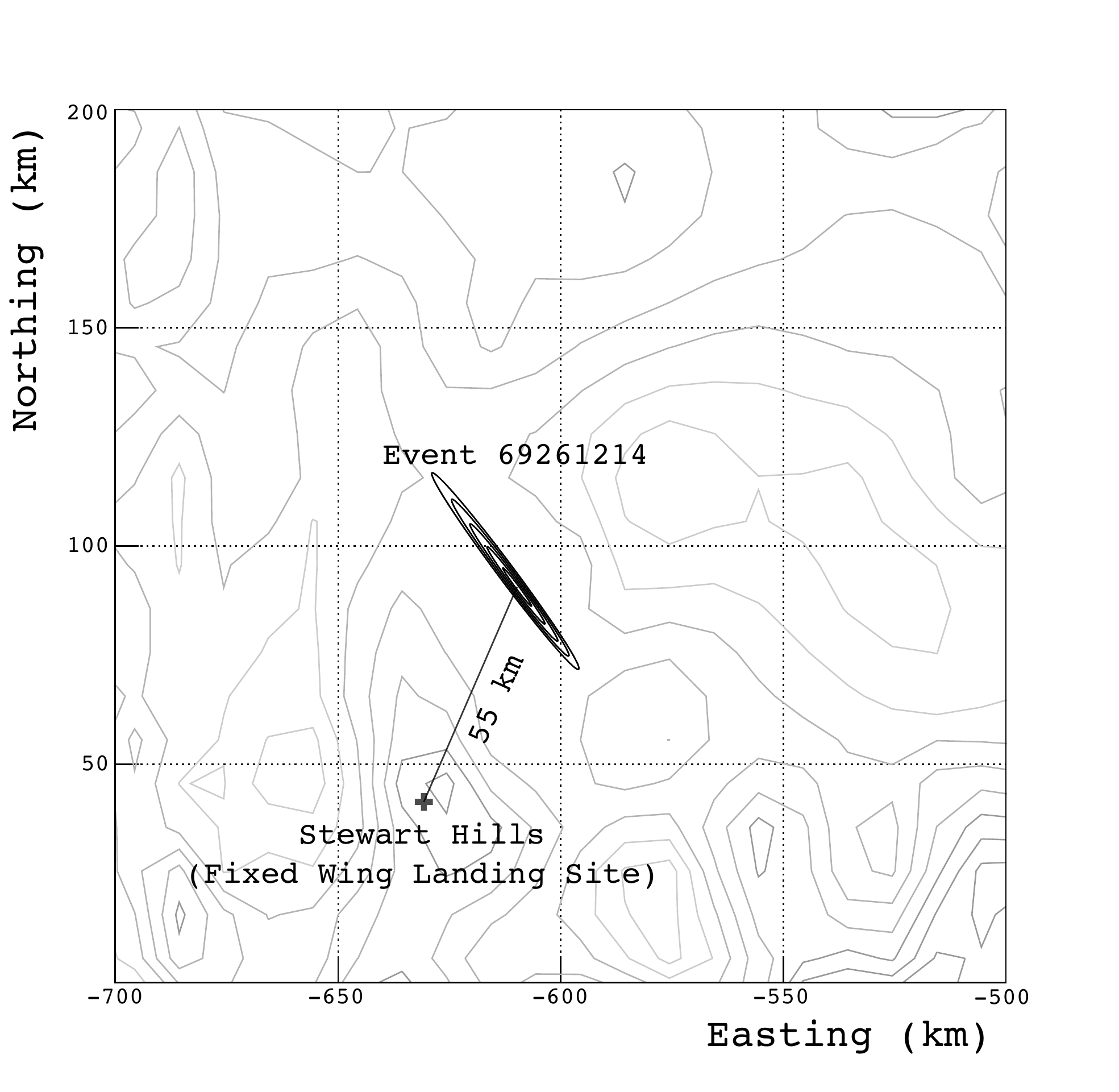}

  \end{minipage}
  \hfill
  \begin{minipage}{.45\textwidth}
    \includegraphics[width=\columnwidth]{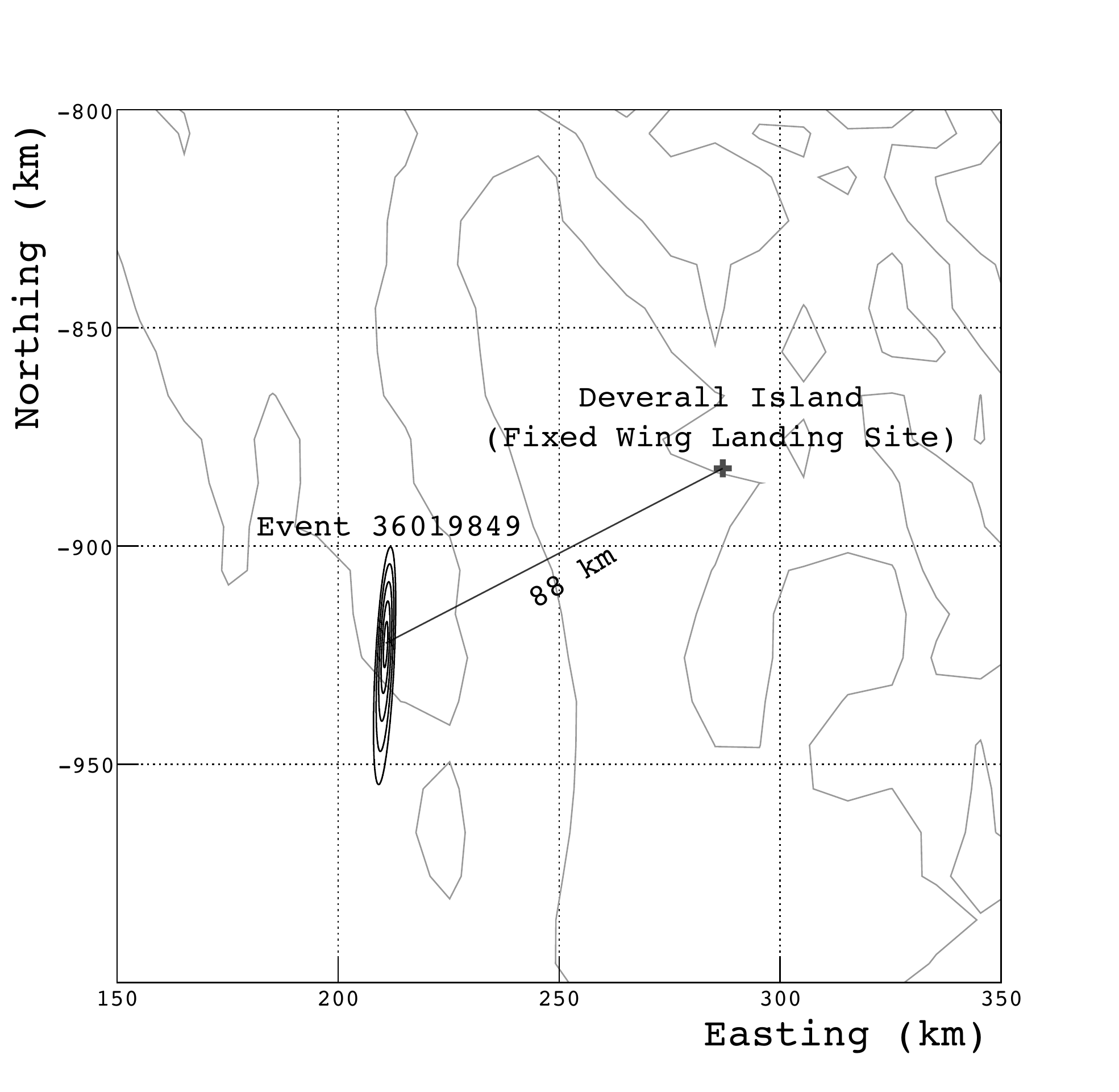}

  \end{minipage} 
  
          \begin{tabular}{|l|l|l|} 
     \hline
     & Event 69261214 & Event 36019849 \\
     \hline
     Reconstructed Event Location & 81.56$^\circ$ W, 84.33$^\circ$ S, 1398 m & 167.13$^\circ$ E, 81.31$^\circ$ S, 2 m\\
     Payload Location &  94.57$^\circ$ W, 85.26$^\circ$ S, 39071 m & 164.47$^\circ$ E, 82.52$^\circ$ S, 39057 m\\
     UTC Time & 2016-12-21 12:20:00 & 2016-12-13 01:53:18\\
     \hline
   \end{tabular}
  
    \caption{Left Panel: Position on the continent for Event 69261214 from Analysis~\textbf{A} with five contours, each representing a one sigma pointing error.  Also shown is the nearest place of known human activity, which is a fixed wing landing site at Stewart Hills, approximately 55~km away.  Right Panel: Position on the continent for Event 36019849 from Analysis~\textbf{B} with five contours, each representing a one sigma pointing error.  The nearest known place of human activity is a fixed wing landing site at Deverall Island, approximately 88~km away.}
  \label{fig:event_maps}
\end{figure*}

The single event in the signal region from Analysis~\textbf{A} (event 69261214), although consistent with the expected number of background events, has a signal shape and polarization consistent with the expected properties of Askaryan emission.  
The emission comes from a location on the continent with deep ice ($\sim$1700~m, roughly 20\% of simulated neutrinos come from ice 1700~m or shallower), and from an angle that is consistent with simulations.  
Shown in Fig.~\ref{fig:event_maps} is the event's position on the continent, along with the nearest known place of human activity, a fixed-wing landing site about 55~km from the event.
We did not record any events from this landing site.
However, we did record five other events within 40~km of this event, including one event within 16~km with a similar signal shape.
There are no other places of known human activity within 100~km of the event.

The single event in the signal region from Analysis~\textbf{B} (event 36019849) is more problematic for interpretation as a neutrino candidate.  
It is consistent with simulated neutrino properties in terms of signal shape, polarization, and exit angle.
However, the emission traces back to the Ross Ice Shelf, where there is only about 300~m of ice before the sea, making neutrino interactions far less likely (only about 1\% of simulated neutrinos come from ice $<=$300~m deep).  
At this location, it is also possible that radio emission due to a neutrino interacting on the ice shelf was first reflected off the water beneath the ice shelf before refracting out of the top surface of the ice. 
  This event is well isolated from other events as well as places of known human activity.  
  Fig.~\ref{fig:event_maps} shows the event on the continent as well as the nearest known place of human activity, a different fixed-wing landing site about 88~km from the event.  
  We did not record any events coming from this landing site.  
  There are no other places of known human activity within 100~km.
  
 We calculate the likelihood ratio of signal over background for each of the events that remain for a variety of observables.  We use a distribution of anthropogenic events that pass all thermal noise cuts as the background sample and a distribution of simulated MC events as signal. Both the simulated signal and background samples are restricted to contain only events with similar signal-to-noise ratios as the candidate event from each analysis.  
  The results of these calculations are shown in Table~\ref{tbl:ll_values}.
  Event 69261214 has a total likelihood ratio of 6.69, meaning that it is more signal-like than background-like.  Event 36019849 has a total likelihood ratio of 0.16, meaning that it is more background-like than signal-like.  
  
  Each of these two events was found by one of the two analyses, but excluded in the other, for reasons we discuss in Section~\ref{sec:summaryEvents}.  
  Event 69261214 was found by Analysis{\bf~A} but not Analysis{\bf~B}, and has a relatively high signal over background likelihood ratio.
  Analysis{\bf~B} rejected this event because there is another event only 16~km away from it.  That nearby event has a very similar signal shape and total likelihood ratio (a likelihood ratio of 5.51), and was recorded within one day of the candidate event.  The distance cut was not used in Analysis{\bf~A}, but such close pairing of events could be indicative of non-neutrino origin.
  
  \begin{table}[t]
\centering

\begin{tabular}{ | l l l | }
\hline
Observable  & Event 69261214 & Event 36019849\\
 & Likelihood Ratio & Likelihood Ratio\\
\hline
Impulsivity  & 1.60 & 0.03\\
Power Gradient  & 1.40 & 2.60\\
Linear Polarization  & 1.15 & 8.40\\
Coherence  & 1.18 & 0.22\\
Hilbert Peak & 2.20 & 1.10\\
\hline
Total Product & 6.69 & 0.16 \\
\hline
\end{tabular}

\caption{Likelihood ratio calculation for the remaining event in each analysis.
The likelihood ratio for signal compared to background is derived using distributions of Monte Carlo neutrino events for the signal distribution and events that passed thermal cuts but failed clustering cuts (consistent with anthropogenic events) as the background distribution.
Observables that go into this calculation are: a measure of how impulsive the signal is (both impulsivity and power gradient), linear polarization fraction, coherence of the signal (measured by the average cross correlation value of individual antenna signals used for reconstruction), and the Hilbert peak of the dedispersed signal.
The total product of all of the likelihood ratios in this table is also shown.
} 
\label{tbl:ll_values}
\end{table}

  In summary, our two independent analyses each found one event consistent with the expected background.
  Combining the limit from the more sensitive of these analyses with previous ANITA limits, we place the strongest constraints on the UHE diffuse neutrino flux at energies between $10^{19.5}$~eV and $10^{23}$~eV .


\section{Acknowledgments} 
ANITA-IV was supported through NASA grant NNX15AC24G.  We are especially grateful to the staff of the Columbia Scientific Balloon Facility for their generous support.  We would like to thank NASA, the National Science Foundation, and those who dedicate their careers to making our science possible in Antarctica.
This work was supported by the Kavli Institute for Cosmological Physics at the University of Chicago.
Computing resources were provided by the Research Computing Center at the University of Chicago and the Ohio Supercomputing Center at The Ohio State University.
A. Connolly would like to thank the National Science Foundation for their support through CAREER award 1255557. 
O. Banerjee and L. Cremonesi's work was supported by collaborative visits funded by the Cosmology and Astroparticle Student and Postdoc Exchange Network (CASPEN).
The University College London group was also supported by the Leverhulme Trust. 
The National Taiwan University group is supported by Taiwan's Ministry of Science and Technology (MOST) under its Vanguard Program 106-2119-M-002-011.
\bibliographystyle{custom_style}
\bibliography{main}

\end{document}